# Can the light be used to treat obesity and diabetes?

Daoudi Rédoane*

*E-mail: redoane.daoudi@unicaen.fr – University of Caen Normandy 14 000 Caen FRANCE

*Abstract.*

*The treatment of obesity and diabetes remains a challenge and the biological mechanisms of these diseases are still not fully understood. Diabetes and obesity are associated with increased risk of the development of cardiovascular complications and there is an urgent need to find novel therapeutic approaches for treating obesity and diabetes. Currently there are several approaches to treat these diseases. Among them chemical uncouplers could be used as an effective treatment for obesity but the dangerous side effects of these compounds has limited their use in vivo. Here we propose a novel theoretical model based on the mechanism of action of chemical uncouplers: the thermogenin-like system (TLS). The TLS may be used in vivo to reproduce the mechanism of action of chemical uncouplers but without their dangerous side effects*.

**Keywords: thermogenesis ; uncoupling ; uncouplers ; weight loss ; obesity ; diabetes ; glucose uptake ; mice ; in vivo ; lipolysis ; luciferase.**

*Highlights:*

+ It may be possible to perform light-induced thermogenesis from most adult tissues *in vivo*, especially when the size of the targeted tissue is large enough to promote an efficient thermogenesis and when there is a large amount of oxygen and mitochondria in the tissue (skeletal muscle for instance).

+ The proposed method (the TLS using a laser) may allow a very precise temporal control of the light-induced thermogenesis *in vivo* compared with cold acclimation or use of chemical uncouplers. In fact, the thermogenesis is stopped when the light is turned off. For this reason, the TLS using a laser may be useful to induce short periods of maximal uncoupling, resulting in an efficient weight loss and glucose uptake without the major side effects of chemical uncouplers that are used at fixed concentrations.

+ The proposed method (the TLS using a laser) may allow a very precise spatial control of the light-induced thermogenesis *in vivo* compared with cold acclimation or use of chemical uncouplers. In fact, the method uses specific promoters of the targeted tissue. Furthermore, the method requires a device implanted under the skin so that it may be possible to target a specific area of the body to induce thermogenesis in this specific region, for instance the thigh muscles. For this reason, the TLS using a laser may be useful to induce a localised thermogenesis *in vivo* in order to reduce side effects. This method may also be used for other research applications.

- The method requires gene therapy.

- The method is invasive.

*Introduction.*

The treatment of obesity remains a challenge and it is reported that increased brown fat activity could be an efficient way to treat obesity by inducing thermogenesis, enhancing metabolism through glucose uptake, oxygen consumption and decreasing adipose tissue triglyceride content through lipolysis. After the recent discovery of the presence of active brown fat in human adults, the study of brown fat has gained interest because its activation could increase weight loss. Additionally, there is growing evidence that white adipocytes can be converted into brown fat-like adipocytes[1], [2] in order to enhance thermogenesis and triglyceride utilization to further promote weight loss[3]. However, the mechanisms involved in this conversion remain unknown in part. For this reason, the amount of active brown fat is insufficient to induce an effective weight loss in human adults. An alternative approach is the use of chemical uncouplers *in vivo* such as FCCP. These compounds induce weight loss, thermogenesis, triglyceride utilization and glucose uptake in white adipocytes[4]. However, the use of chemical uncouplers *in vivo* is limited since these compounds have significant side effects like hyperthermia, tachycardia, diaphoresis, eventually leading to death[5]. Another study shows that certain chemical uncouplers have a dynamic range and can be safely used in vivo because they induce a limited uncoupling[6]. Nevertheless, these compounds may have limited effects on weight loss. Here we describe a novel theoretical method to induce an autoregulated artificial uncoupling *in vivo*, potentially resulting in safe and effective thermogenesis, glucose uptake, lipolysis and subsequent weight loss: a thermogenin-like system (TLS). Because the TLS may induce thermogenesis, glucose uptake, oxygen consumption and triglyceride utilization from most adult tissues, it may be used as an interesting alternative treatment for obesity. Because glucose uptake reduces blood glucose levels, the TLS could decrease insulin production and subsequent lipogenesis, suggesting that the TLS could constitute an interesting approach to the treatment of obesity and diabetes. Thus, we propose the TLS as an alternative treatment for obesity *in vivo*. Indeed, the fact that the TLS requires ATP is remarkable, the system will stop if the level of ATP dramatically decreases, preventing both the uncoupling-induced proton leak and the cell death or cell dysfunction *in vivo*. Moreover, the TLS is designed to function in a specific tissue and not at the whole-organism level in contrast to an intravenous injection of chemical uncouplers. For example, the TLS may be used in white adipocytes to induce thermogenesis from these cells. The system may also be used in endothelial cells because there are large numbers of endothelial cells in an adult organism and there is a large amount of oxygen in this tissue, allowing for efficient thermogenesis from these cells. We think that the TLS could be used in combination with approaches that increase brown fat activity to treat obesity but further studies are needed.

*Thermogenin.*

Another way to promote uncoupling *in vivo* is the thermogenin (also called UCP1) system. This protein is naturally present in the mitochondria of brown adipose tissue and it is used to produce heat by non-shivering thermogenesis. UCP1 is activated by fatty acids in the brown fat in mammals. Therefore, it is not possible to express UCP1 in other cells in order to induce an effective thermogenesis since most cells may not possess the machinery required for fatty acids uptake and the molecular mechanisms underlying fatty acids transport remain elusive in part. Furthermore, even if white adipocytes possess a large amount of fatty acids, this system would be ineffective because it is necessary to induce lipolysis to stimulate UCP1 and to promote uncoupling. Lipolysis is induced by a sympathetic stimulation *in vivo* in the brown fat and white adipocytes don't possess an adequate sympathetic innervation *in vivo* to support an effective thermogenesis *in vivo*. Although an artificial injection of norepinephrine could be used to stimulate engineered white adipocytes *in vivo* (UCP1 and norepinephrine-lipolysis pathways double positives fat cells), it could induce major side effects like tachycardia and death. For these reasons it is possible to establish a theoretical experimental model based on the thermogenin system: the TLS (or UCP1-like system).

### *The thermogenin-like system: TLS.*

The first step of this model is the *in vivo* transduction of adipose tissue with adeno-associated viral (AAV) vectors. The following transgenic DNA constructs are used *in vivo*: the light-driven inward H+ pump PoXeR targeted to the inner mitochondrial membrane, firefly luciferase, luciferin-regenerating enzyme (LRE), cysteine racemase and thioesterase. Now, we describe the function of each actor. PoXeR is a natural light-driven inward proton pump found in *Parvularcula oceani*, a deep-ocean marine bacterium. This pump controls the unusual directionality opposite to normal proton pumps and may lower the proton motive force when expressed in the inner mitochondrial membrane, producing heat similarly to thermogenin. A study shows that PoXeR can be expressed in mouse neural cells and it is functional in these cells[7]. Another study shows that it is possible to construct photoenergetic mitochondria in cultured mammalian cells expressing another light-driven proton pump derived from *Haloterrigena turkmenica*[8]. Taken together, these results suggest that PoXeR may be a good candidate for the TLS *in vivo* in mammals. Apart from PoXeR, another protein may be used for the TLS, this is the *Anabaena* sensory rhodopsin (ASR). This photochromic sensor exhibits an inward proton transport activity driven by absorption of a single photon when Asp217 is replaced by Glu (D217E)[9]. Similar to PoXeR, D217E ASR may be a good candidate for the TLS but in contrast to PoXeR, its functionality in mammalian cells remains unknown. Note that we use a pump because channelrhodopsins exhibit a strong light-dependent inactivation[10] and are not suitable for long-term induction of proton leak across the inner mitochondrial membrane. Because PoXeR is a light-driven inward proton pump, the light is required for its activation. Firefly luciferase is a light-emitting enzyme responsible for the bioluminescence of fireflies. It can be expressed in adipose tissue and more generally in cells since it is largely used as a reporter gene. Firefly luciferase catalyses the oxidation of firefly luciferin and requires oxygen and ATP. There is an overlap between the emission spectrum of firefly luciferase and the absorption spectrum of PoXeR or D217E ASR so that the light produced by firefly luciferase can activate PoXeR and induces the proton leak, generating heat. Because the proton leak decreases ATP production, cell death may rapidly occur. However, note that firefly luciferase requires ATP and it is interesting to note that the reaction will stop if the level of ATP dramatically decreases, thus preventing both the proton leak and the cell death. Therefore, the TLS may be used in normal cells to promote uncoupling-induced weight loss without significant adverse effects. It may be possible to express PoXeR (and other transgenes of the TLS) under the control of a weak promoter to reduce the functionality of the system and subsequent weight loss. It may be also possible to express PoXeR (and other transgenes) under the control of a strong promoter to enhance weight loss without significant side effects. In addition, it is interesting to obtain a spatial control of the system in order to restrict transgenes expression to the adipose tissue. For this reason, we propose to use the following transgenic DNA constructs *in vivo*[11]:

   [Adiponectin promoter]-PoXeR targeted to the inner mitochondrial membrane

   [Adiponectin promoter]-firefly luciferase

   [Adiponectin promoter]-LRE

   [Adiponectin promoter]-cysteine racemase

   [Adiponectin promoter]-thioesterase

Because the bioluminescence of firefly luciferase decreases over time, it is necessary either to continually bring the substrate (i.e firefly luciferin) by continuous infusion and/or to use a system that regenerates the firefly luciferin from oxyluciferin, the product of the reaction. A study shows that oxyluciferin is enzymatically regenerated into firefly luciferin by LRE *in vitro*[12]. This system can be improvable by using cysteine racemase and thioesterase[13]. Finally, it is possible to concentrate the substrate firefly luciferin in adipose tissue *in vivo* by using a suitable linker (firefly luciferin prodrug). Then, mice are injected with a firefly luciferin or a firefly luciferin prodrug so that the active firefly luciferin is mainly released in adipose tissue.

Now, the TLS needs to be sensitized to the light. In fact, all trans retinal is required for light-induced PoXeR activation *in vivo*. All trans retinal may be not naturally present in adipose tissue, thus it is necessary to administer it by continuous infusion, it is not enzymatically regenerated in contrast to firefly luciferin. Because all trans retinal is hydrophobic it can be encapsulated in liposomes to reach the adipose tissue *in vivo*. Hydrophobic compounds have affinity to the phospholipid bilayer of liposomes.

Finally, few hours after all trans retinal injection, mice are injected with a firefly luciferin or a firefly luciferin prodrug.

Therefore, the TLS may be used for treating diabetes and obesity *in vivo* by inducing a spatially and temporally controlled thermogenesis, glucose uptake and lipolysis *in vivo* in the adipose tissue. Furthermore, it may be possible to induce a moderate or an elevated thermogenesis with this system if an adequate promoter is used.

Now, we propose an experimental procedure in order to check the functionality of the TLS *in vivo* in mice. Firstly, use the previous transgenic DNA constructs for the *in vivo* transduction of adipose tissue with AAV vectors. In fact, a study indicates that AAV-mediated genetic engineering of white and brown adipose tissue is possible in adult mice[14]. Then, use a continuous infusion of all trans retinal in mice. Few hours after all trans retinal injection, use a continuous infusion of firefly luciferin or firefly luciferin prodrug. Finally, estimate the weight loss, blood glucose levels or measure body temperature of mice using infrared imaging. Indeed, a study shows that it is possible to measure brown adipose tissue thermogenesis using infrared imaging in mice[15]. Moreover, it may be possible to detect an increased temperature *in vivo* if the TLS is functional since a study shows that the elevated temperature originates from a collective effect[16]. A tissue such as the adipose tissue containing N cells of individual size L and individually delivering a heat power P has a size $N^{1/3}L$ and delivers a heat power PN. Thus, the increased temperature ΔT is defined as:

$$\Delta T = \frac{P}{kL} N^{\frac{2}{3}}$$    k is the thermal conductivity.

For this reason, it is not possible to check the functionality of the TLS *in vitro* and/or in single living cells and we speculate that an effective thermogenesis-induced weight loss could be obtained *in vivo* because the size of the adipose tissue is large enough to promote a significant thermogenesis.

### *Limitations of the TLS and conclusion.*

To conclude, the TLS remains a theoretical model that requires further studies. The proposed system may have several advantages: it may be used for treating obesity and diabetes *in vivo* by inducing a spatially and temporally controlled uncoupling-induced thermogenesis *in vivo*. The TLS may be used to promote an effective thermogenesis-induced weight loss *in vivo* without significant side effects compared to chemical uncouplers. In fact, the TLS requires ATP and the system will stop if the level of ATP dramatically decreases, preventing both the uncoupling-induced proton leak and the cell death or cell dysfunction *in vivo*. Thus, the TLS using firefly luciferase may dynamically adapt to the ATP level of adipocytes to prevent cell death *in vivo*, thermogenesis-related and ATP depletion-related side effects (1) and to promote uncoupling mainly when energy expenditure is necessary (i.e high ATP levels) (2) **(figure 1)**.

However, the TLS also suffers from various drawbacks. Firstly, considerable work is needed until gene therapy can be used for safe treatment of obesity and diabetes in humans. Furthermore, it is not clear whether all trans retinal has a significant toxicity when injected *in vivo* as previously described. It would be possible to use either a moderate dose of all trans retinal (but sufficient to allow for efficient thermogenesis in white adipocytes) or a high dose of all trans retinal targeted to the adipose tissue via all trans retinal prodrug. Then, the quantum efficiency of firefly luciferase is high so that 0.88 photon is emitted per consumed ATP molecule[17]. Considering that $10^9$ molecules of ATP are in solution in a cell at any instant[18], 880 000 000 photons can be emitted at any instant in a cell. The number of D217E ASR proteins in the inner mitochondrial membrane per cell is likely close to the number of adiponectin proteins per cell since these two proteins are controlled by the same promoter (the differences in the turnover rates are omitted). Thus, the number of D217E ASR proteins per cell is expected to be smaller than the number of photons emitted at any instant. In fact, a recent study estimates that a yeast cell contains about $4.2 \times 10^7$ proteins[19]. The number of proteins in an adipocyte is probably around this value. For this reason, a maximal uncoupling could occur at any instant in an adipocyte, even if ATP levels are dramatically decreased, resulting in cell death (because the number of ATP molecules and consequently the number of photons emitted is still greater than the number of D217E ASR proteins). It is important to remember that one photon is sufficient to activate D217E ASR. This issue can be encountered by using another luciferase with a quantum efficiency smaller than the quantum efficiency of firefly luciferase. For instance, the bacterial luciferase enzyme requires 60 ATP per photon emitted (490nm)[20]. However, this luciferase uses n-decanal as a substrate and it is toxic to cells. Thus, the TLS should be used with another luciferase with a very low quantum efficiency, with a substrate that is non-toxic to cells and with an overlap between the emission spectrum of this luciferase and the absorption spectrum of PoXeR or D217E ASR. Because the bioluminescence quantum yield can be enhanced[21], we hypothesize that a luciferase with a very low quantum efficiency could be designed and used as a tool for the TLS.

An alternative method is to use a laser as the source of light instead of a luciferase system. The excitation wavelength should be 567nm that is the wavelength of maximum absorbance of PoXeR and D217E ASR. This method is useful in order to simplify the TLS. By using a laser as the source of light, it is not necessary to find a luciferase with the previous characteristics. Furthermore, a maximal uncoupling could occur in an adipocyte when the light is turned on because the number of photons emitted by the laser at any instant is greater than the number of D217E ASR proteins. The maximal uncoupling is useful to induce an effective weight loss. However, because a maximal uncoupling is expected when the light is turned on and that the system is not autoregulated compared to the TLS using a luciferase, it is necessary to turned off the light when ATP levels are dramatically decreased. Thus, the optimal duration of a light pulse and the optimal time gap between two light pulses should be determined *in vivo* to automatically turned off the light when ATP levels are dramatically decreased and to obtain a maximal uncoupling and an effective weight loss without side effects. Then, it is necessary that the light reaches the adipocytes *in vivo* to stimulate PoXeR or D217E ASR. For this reason, the TLS using a laser is an invasive method because the device should be implanted under the skin to allow the light to reach the adipocytes *in vivo*. We propose the TLS using a laser as a potential future method to induce an effective weight loss and glucose uptake without major side effects compared to chemical uncouplers. This method is derived from optogenetics **(figure 2)**.

**Figure 1. The TLS (or UCP1-like sytem)** *in vivo* **using a luciferase.**

**A.** Experimental procedure. The TLS may be used for treating obesity and diabetes in vivo by inducing an effective thermogenesis-induced weight loss or glucose uptake *in vivo* if the size of the target tissue is large enough to produce significant local heat *in vivo*. The local heat production can be measured by using infrared imaging for adipose tissue.

**B.** Detailed pathways involved in local heat production using the TLS. The cysteine racemase, not shown here, is used to produce D-cysteine from L-cysteine *in vivo*.

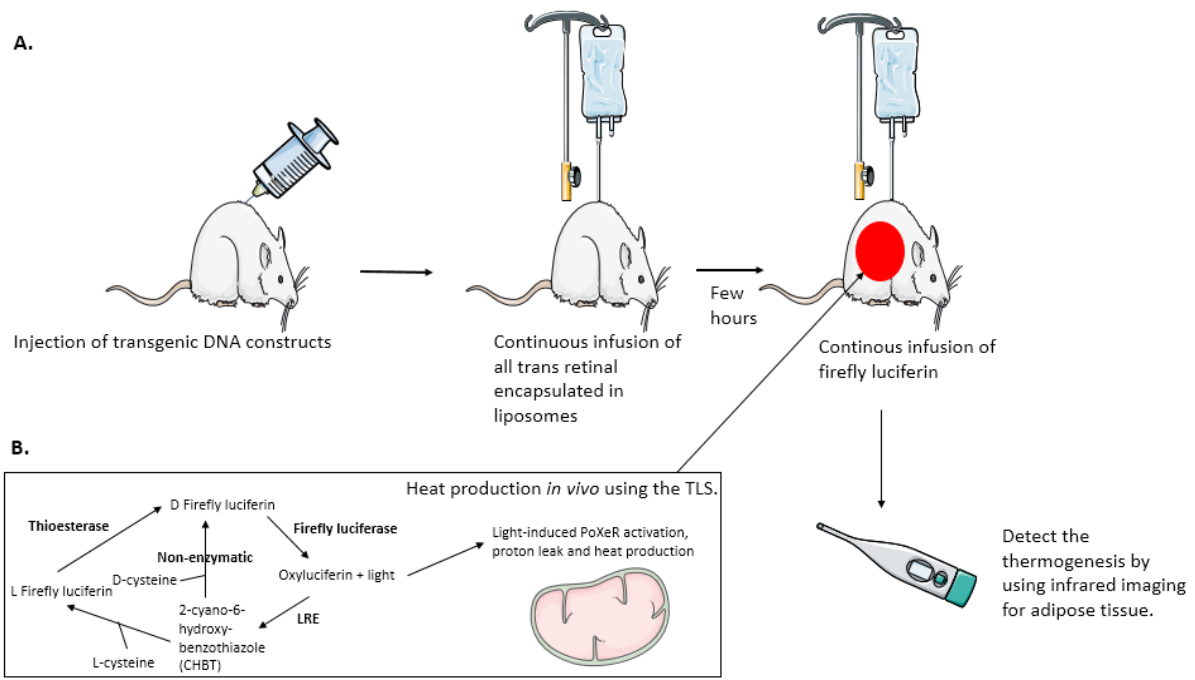

**Figure 2. The TLS (or UCP1-like sytem)** *in vivo* **using a laser.**

**A.** Theoretical predictions of thermogenesis and glucose uptake rates variations using the TLS with a laser. In the theoretical model it is expected that the light induces a maximal uncoupling, resulting in increased thermogenesis and glucose uptake. When the light is automatically turned off, the uncoupling is stopped and the thermogenesis and glucose uptake rates are decreased.

**B.** Theoretical predictions of ATP level variations using the TLS with a laser. In the theoretical model it is expected that the light induces a maximal uncoupling, resulting in decreased ATP level. When the light is automatically turned off, the uncoupling is stopped and the ATP level is increased.

**C.** Experimental procedure. Firstly, use mice injected with transgenic DNA constructs (PoXeR or D217E ASR targeted to the inner mitochondrial membrane) and use a continuous infusion of all trans retinal encapsulated in liposomes (not shown here). Then, a laser delivers light pulses (567 nm). The light should be automatically turned off when ATP levels are dramatically decreased and automatically turned on when ATP levels are sufficient for cell survival to increase thermogenesis and glucose uptake. Then, it is necessary to determine *in vivo* the duration of a single light pulse and the time gap between two light pulses to optimize the system. Like the TLS using a luciferase, the TLS using a laser may be used for treating obesity and diabetes *in vivo* by inducing an effective thermogenesis-induced weight loss or glucose uptake *in vivo* if the size of the target tissue is large enough to produce significant local heat *in vivo*. The local heat production can be measured by using infrared imaging for adipose tissue. The TLS using a laser may be more efficient than the TLS using a luciferase and without major side effects.

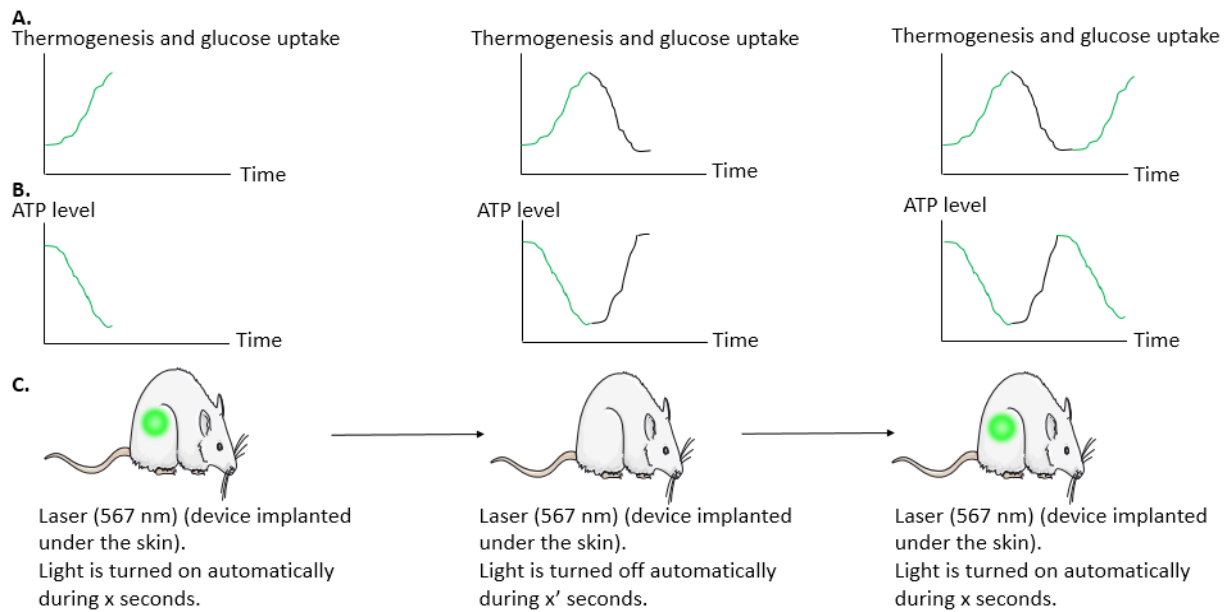

*Acknowledgement.*

I would like to thank Professor Alan Rendall for agreeing to endorse me.